\documentclass[twocolumn,superscriptaddress,prl]{revtex4}
\usepackage{amssymb}
\usepackage{graphicx}
\usepackage{dcolumn}
\usepackage{bm}
\usepackage{amsmath}

\newcommand{\ket}[1]{| #1 \rangle}

\begin{document}

\title{Topological Stability of Stored Optical Vortices}
\author{R. Pugatch}
\affiliation{Department of Physics of Complex Systems, Weizmann
Institute of Science, Rehovot 76100, Israel}
\author{M. Shuker}
\affiliation{Department of Physics, Technion-Israel Institute of
Technology, Haifa 32000, Israel}
\author{O. Firstenberg}
\affiliation{Department of Physics, Technion-Israel Institute of
Technology, Haifa 32000, Israel}
\author{A. Ron}
\affiliation{Department of Physics, Technion-Israel Institute of
Technology, Haifa 32000, Israel}
\author{N. Davidson}
\affiliation{Department of Physics of Complex Systems, Weizmann
Institute of Science, Rehovot 76100, Israel}

\begin{abstract}
We report an experiment in which an optical vortex is stored in a
vapor of Rb atoms. Due to its $2\pi$ phase twist,  this mode, also
known as the Laguerre-Gauss mode, is topologically stable and
cannot unwind even under conditions of strong diffusion. For
comparison we stored a Gaussian beam with a dark center and a
uniform phase. Contrary to the optical vortex, which stays stable
for over $100 \mu s$, the dark center in the retrieved flat-phased
image was filled with light after storage time as short as $10 \mu
s$. The experiment proves that higher electromagnetic modes can be
converted into atomic coherences, and that modes with phase
singularities are robust to decoherence effects such as diffusion.
This opens the possibility to more elaborate schemes for two
dimensional information storage in atomic vapors.
\end{abstract}

\maketitle

 Topological defects are abundant in diverse coherent media,
ranging from neutron stars \cite{neutronstar1,neutronstar2},
superconductors \cite{SuperfluidityBook}, superfluid $^{4}$He and
$^{3}$He \cite{SuperfluidityBook,QuantizedVorticesBook},
Bose-Einstein condensates \cite{BECbookpethicksmith}, effective
two dimensional Rb atoms \cite{dalibardBKT} and light
\cite{PadgettLG}. A topological defect is a spatial configuration
of a vector field that defines a distinct homotopy
class \cite{MerminTopologicalDefects}, it is therefore stable
against continuous deformations which cannot cause it to
decay or to \textquotedblleft unwind\textquotedblright.
It has been suggested that this robustness can be harnessed to
combat decoherence effects by using topological defects as qubits
in a quantum computation scheme
\cite{QuantumComputingTopology,Kitaev}.

 Perhaps the simplest example of a topological defect is a vortex. A vortex in a coherent media
occurs whenever there is a $2\pi $ dislocation of the phase around
a given point in space
\cite{MerminTopologicalDefects,BECbookpethicksmith,QuantizedVorticesBook}.
In particular, a light vortex is an electromagnetic mode with a
phase $e^{i\theta \left( x,y\right) }$, where $\theta \left(
x,y\right) =\arctan \left( y/x\right) $, is defined in the plane
perpendicular to the propagation direction $\hat{z}$. There exist
a family of free space solutions of Maxwell's equations, known as
the Laguerre-Gauss modes, that have this phase factor
\cite{Allen92,PadgettLG}.

 In this work we report an experiment in which a
light vortex was stored in \textquotedblleft
hot\textquotedblright\ Rb vapor and then retrieved. We show that
due to its topological stability, the retrieved light beam
maintains its phase singularity, thus staying dark at the center
in a regime of strong diffusion for storage times up to $110$ $\mu
s$. For comparison, we stored a Gaussian beam with a uniform phase
and a dark center, and observed that the center was filled with
light after storage of only $10$ $\mu s$ due to diffusion of atoms
to the center. We suggest to use this topological robustness for
improved storage of two dimensional information e.g. images. The
prospect of creating more elaborate defects is also briefly
discussed.

\begin{figure}[ht]
\begin{center}
\includegraphics[width=8.5cm]{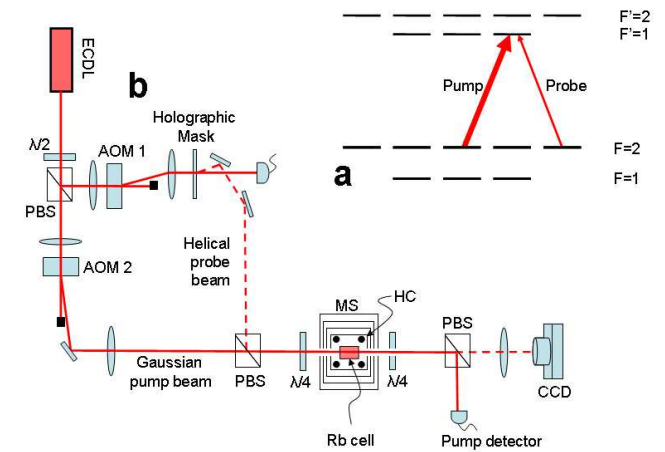}
\end{center}
\par
\caption{(color on-line) Experimental setup for optical vortex storing. a. The energy level scheme of the $D1$ transition of $^{87}Rb$ showing
the three levels of the $\Lambda$-system and the pump and probe
transitions. b. The experimental setup. ECDL - external cavity
diode laser, PBS - polarizing beam splitter, AOM - acousto-optic
modulator, $\lambda$/2 - half wave-plate, $\lambda$/4 - quarter
wave-plate, MS - magnetic shields, HC - Helmholtz coils.}
\label{fig:setup}
\end{figure}

 When storing light in an electromagnetically induced transparency
(EIT) medium, the information contained in a weak ``probe'' field,
namely its amplitude and phase, is continuously transformed onto
the atomic coherence
\cite{LukinDarkStatePolaritons1,LukinDarkStatePolaritons2}. The
propagation of the combined light-matter excitation in the medium
(``dark-state polariton''), its deceleration down to zero velocity
for a controlled period of time (storage), and the subsequent
retrieval of the probe pulse, are controlled by switching a strong
``pump'' field. In the simplest case, this coherent atomic medium
is described as a three-level $\Lambda -$system with the probe and
pump fields resonant with the two optical transitions
$1,2\leftrightarrow 3$ (see Fig. \ref{fig:setup}.a). In a density
matrix formalism the dark-state is identified with the ground
level coherence $\rho
_{12}\left( \mathbf{r},t\right) =\left( -g/\Omega \right) E\left( \mathbf{r}%
,t\right) $, where $g$ is the atom-field coupling constant,
$\Omega$ is the Rabi frequency of the pump and $E(\mathbf{r},t)$
is the slowly varying amplitude of the probe field
\cite{LukinDarkStatePolaritons1}. Assuming the pump field has a constant phase relative to the probe, we note that the atomic coherence follows the complex probe field $E(\mathbf{r},t)$, and hence for a general
transverse electromagnetic mode it may be different for each point
on the transverse plane.

 A particularly interesting electromagnetic mode is the so called Laguerre-Gauss (LG) mode, also known as an optical vortex or helical beam.  The electric field of the LG$_{0}^{m}$ mode is given in polar coordinates (in the $z=0$ plane) by \cite{Siegman}
$E_{m}^{\text{helical}}\left(r,\theta\right) =A_{m}\left( r,w_{0}\right) e^{-im\theta }$ where $w_{0}$ is the waist, $m$ is the winding number and $A_{m}\left( r,w_{0}\right) =\frac{1}{w_{0}}\sqrt{\frac{2P}{\pi m!%
}}\left( \sqrt{2}r/w_{0}\right) ^{m}\exp \left(
-r^{2}/w_{0}^{2}\right)$ is the ``ring shaped'' radial
cross-section with $P$ the total intensity. The LG modes have some
unique properties, e.g. carrying orbital angular momentum
\cite{Allen92,PadgettLG}. The aspect relevant for this work is
that this mode has a $2\pi m$ phase twist around its dark center.
The dark center of the light vortex --- a result of the phase
singularity --- can be intuitively understood for the $m=1$ case,
as a destructive interference between any two diametrically
opposed (antipodal) pair of points around the center, since any
such pair has a $\pi$ phase shift. The winding number $m$ is a
topological invariant characterizing this $U(1)$ defect, which
cannot change under continuous deformations of the probe field.
This suggests that by storing a light vortex in the atomic
coherences, we can create a coherence field which is robust to
diffusion, since diffusion is, on average, a continuous process.

 To account for the effect of diffusion on the retrieved signal, we
assume that the internal state of each individual atom does not
change as a result of diffusion, and so it carries with it the
stored complex amplitude. The ensemble average of the different
atoms arriving to the same small macroscopic volume will determine
the retrieved probe field \cite{RamseyDiffusion}. Therefore the
atomic diffusive motion for a large number of atoms can be
described as a continuous diffusion process of the density matrix
field, by adding the term $\dot{\rho}_{\text{diff}}=D\nabla
^{2}\rho $ to the Bloch equations, where $D$ is the diffusion
coefficient. Eventually, the coherence field after a storage time
$t$, and neglecting all other decay mechanisms, is calculated by
propagating the initial coherence forward in time, using the three-dimensional diffusion propagator,
$G(\mathbf{r,r}^{\prime},t)=(2\pi Dt)^{-3/2}\exp [-\left(
\mathbf{r-r}^{\prime }\right) ^{2}/\left( 4Dt\right) ]$. The
retrieved probe field, $E_{\text{ret}}$, is given by
$E_{\text{ret}}(\mathbf{r},t)=(-\Omega/g)\rho_{12}(\mathbf{r},t)$.

 Taking the initially stored coherence field to be $\rho_{12}\left(
r,\theta ,t=0\right) =\left( -g/\Omega \right)
E_{m}^{\text{helical}}\left( r,\theta \right) $, the expected
field after diffusion of duration $t$ is a scaled LG mode,
\begin{equation}\label{eq:Helical}
\rho _{12}^{\text{helical}}\left( r,\theta ,t\right) =\frac{\left(
-g/\Omega \right) }{\sqrt{s(t)^{m+1}}}A_{m}\left(
r,\sqrt{s(t)}w_{0}\right) e^{-im\theta },
\end{equation}
where $s\left( t\right) =\left( {w}_{0}^{2}+4Dt\right) /w_{0}^{2}$
is a scaling factor. 
Note that at the center $\rho _{12}\left( r=0,t\right) =0$ for all
$t$. For the sake of comparison, we examine the diffusion of a
Gaussian beam with a dark hole of radius $r_{0}$ at the
center, and a \textquotedblleft flat\textquotedblright\ (constant) phase, $%
E^{\text{flat}}\left( r\right) =\Theta \left( r-r_{0}\right)
A_{0}\left( r,w_{0}\right)$, where $\Theta \left( r\right) $ is
the step function. The field of atomic coherence after time $t$ for this beam profile is given by%
\begin{equation}\label{eq:Flat}
\rho _{12}^{\text{flat}}\left( r,t\right) =\frac{\left( -g/\Omega \right) }{%
2Dt}\sqrt{\frac{2P}{\pi w_{0}^{2}}}\int_{r_{0}}^{\infty
}dr^{\prime }r^{\prime }e^{-\frac{r^{2}+sr^{\prime
2}}{4Dt}}I_{0}\left( \frac{rr^{\prime }}{2Dt}\right) ,
\end{equation}%
where $I_{0}\left( x\right) $ is the modified Bessel function of
order zero \cite{AbramowitzStegun}. Taking $r_0=w_0/2$ the
coherence at the center ($r=0$) is proportional to $\exp [1/\left(
1-s\right)/4 ]/s$, so the hole is expected to be filled at
$t\approx 0.15w_{0}^{2}/D$.

 In order to store the light in this experiment, we employed the EIT within the $D1$
transition of $^{87}$Rb \cite{LukinStorageOfLight}. The energy
level scheme is presented in Fig. \ref{fig:setup}.a, showing the
$\Lambda$-system and the pump and probe transitions. The
experimental apparatus is depicted in Fig. \ref{fig:setup}.b. An
external-cavity diode-laser (ECDL) was locked to the
$F=2\rightarrow F'=1$ transition. The laser was divided into two
beams of perpendicular linear polarizations, the pump and the
probe, using a polarizing beam-splitter (PBS). The pump beam was
passed through an acousto-optic modulator (AOM) and the first
order was used for the experiment, allowing us to control both the
frequency and the intensity of the pump. The pump was shaped as a
Gaussian beam with a waist of $w_{\text{pump}}=2.2 mm$ (at the
center of the vapor cell) and a total intensity of $1.1 mW$. The
probe channel included a second AOM in a similar arrangement, as
well as an off-axis holographic binary mask, used to shape the
probe as a helical beam with $m=1$. The waist of the helical probe
was set to $w_0=670$ $\mu m$ at the center of the vapor cell and
its intensity was $32$ $\mu W$. The small size of the probe
insured that it will experience a nearly constant pump intensity
and phase, and also guaranteed the dominance of the diffusion
mechanism over other decoherence mechanisms. The pump and the
probe beams were recombined on a second PBS, and co-propagated
toward the vapor cell. A quarter wave-plate before the cell
converted the pump and the probe from linear to $\sigma^{+},
\sigma^{-}$ polarizations, suitable to the transitions presented
in Fig. \ref{fig:setup}.a. A $5$ cm long vapor cell containing
isotopically pure $^{87}$Rb and $10$ Torr of Neon buffer gas was
used. The temperature of the cell was set to $\sim65^{\circ}$ C,
providing a Rubidium vapor density of $\sim4\times10^{11} /cc$.
The cell was placed inside a four-layered magnetic shield, and a
set of Helmholtz coils allowed us to determine the axial magnetic
field. After the beams passed through the vapor cell they were
separated using polarization optics. The pump beam was measured by
a photo-diode detector and the probe beam was imaged onto a CCD
camera. We used a small, $B_z=50$ mG axial magnetic field to set
the quantization axis. A suitable frequency shift between the pump
and the probe was introduced by scanning the EIT resonance and
setting the pump AOM frequency to its center.

 A typical experimental sequence started by applying the pump beam
for a long duration, optically pumping a substantial atomic
population to the $\ket{F=2,m_F=+2}$ state. A Gaussian probe pulse
with $\sigma=30$ $\mu s$ was then sent into the cell, and was
slowed to a group velocity of $\sim 1000$ $m/s$ (delay of $\sim
50$ $\mu s$). During the passage of the probe pulse in the vapor
cell, the pump beam was turned off, storing the probe beam in the
atomic ground-state coherence. After a certain storage duration
(during which diffusion of the atoms occurred) the pump beam was
turned back on, restoring the probe, which was then imaged onto a
CCD camera. The CCD camera was triggered to measure \emph{only}
the restored part of the probe (to reduce the noise, we typically
averaged $128$ single shot images). The effect of atomic diffusion
on the spatial shape of the restored light beam was studied by
measuring it for different storage durations.

\begin{figure}[ht]
\begin{center}
\includegraphics[width=8.5cm]{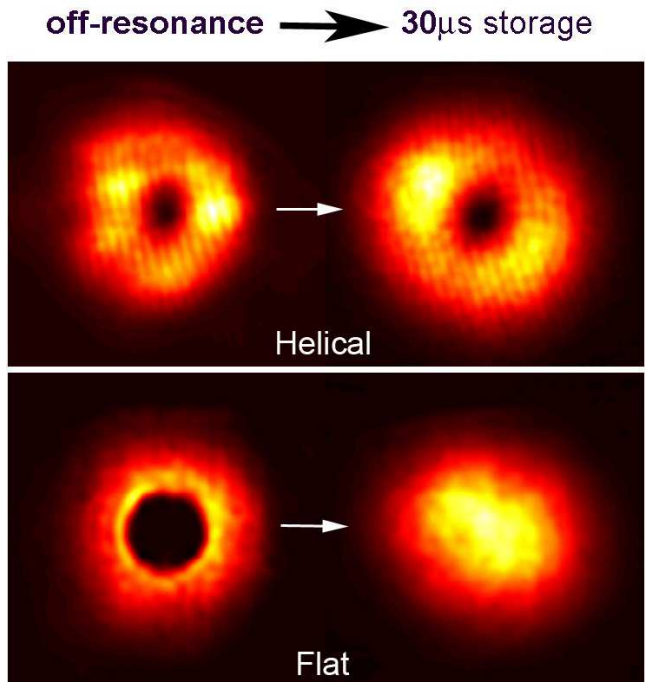}
\end{center}
\par
\caption{(color on-line) The effect of diffusion on a helical beam
and a flat phase beam. The left column shows the initial shape of
a helical (top) and a flat phase (bottom) beams before any
diffusion (off-resonance). The right column shows both beams after
storage of $30$ $\mu s$. The center remains dark in the helical
beam, while completely filling with light in the flat phase beam.}
\label{fig:NicePicture}
\end{figure}

 In the main experiment we studied the effect of diffusion on the spatial shape of a helical
beam with $m=1$. In a control experiment we used a transverse
Gaussian beam ($\text{TEM}_{00}$ mode, $w_{0}=670 \mu m$) with a
darkened center that was created by using a small circular
beam-stop, and which we imaged into the cell center. We also
verified that the dark center of the beam persists throughout the
length of the cell, by first measuring it along a parallel optical
path of equal length outside the magnetic shield. The radius of
the circular stop was $r_0=300$ $\mu m$. While the intensity
pattern (lower-left image in Fig. \ref{fig:NicePicture}) is also
ring-shaped and similar to that of helical beam (upper-left image
in Fig. \ref{fig:NicePicture}), the partially-blocked Gaussian
beam had a flat phase, as opposed to the twisted phase of the
helical beam. Note that the actual radius of the dark center of
the flat phase beam when imaged on the atoms was $\sim 400 \mu m$,
since we used an imaging system with a magnification of $M=1.4$.

 The main result of this work is depicted in Fig.
\ref{fig:NicePicture}. The left column shows the two beam-shapes
we tested before any interaction with the atoms (the images were
taken by detuning the laser from the atomic resonance). The right
column shows the same beams after $30$ $\mu s$ of storage in the atomic vapor, when substantial diffusion took place.
It is evident that while the central hole in the flat phase beam is
completely filled, the central hole of the helical beam remains
dark.

\begin{figure}[ht]
\begin{center}
\includegraphics[width=8.3cm]{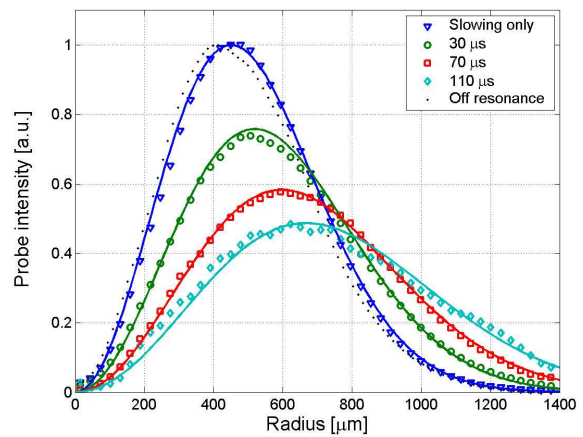}
\end{center}
\par
\caption{(color on-line) Averaged radial cross-sections of the
restored helical beam for different storage durations ($30$ $\mu s$,
$70$ $\mu s$ and $110$ $\mu s$). The cross-sections of the
off-resonance and slowing cases are also shown (the small
difference between them is mainly due to diffusion during the
slowing). It is evident that the center remains completely dark,
even for the longest storage duration. The theoretical prediction
(solid lines) are in excellent agreement to the experimental
results (symbols). As explained in the text, the effect of
diffusion on this transverse mode is to increase its waist $w_0$
to $\sqrt{w_0^2+4Dt}$, where $D = 11$ $cm^2/s$.}
\label{fig:HelicalRadialDistribution}
\end{figure}

\begin{figure}[h]
\begin{center}
\includegraphics[width=8.3cm]{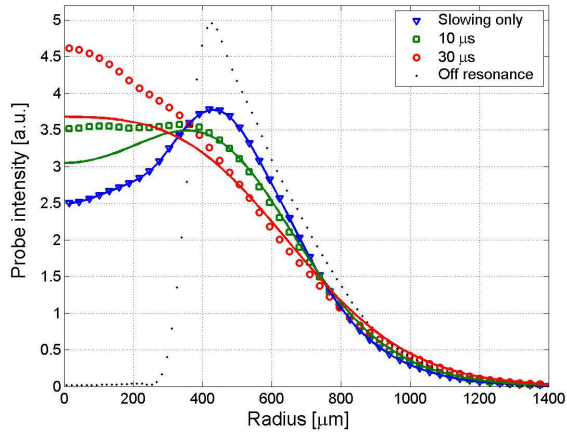}
\end{center}
\vspace{0.4cm} \caption{Averaged radial cross-sections of the
restored flat phase beam with a dark center for two storage
durations ($10$ $\mu s$, $30$ $\mu s$). Also shown are the slowed
and off-resonance cross-sections. Due to the diffusion the dark
center is already filled when the beam is only slowed. The
theoretical prediction (solid lines) capture this essential
phenomenon, but is less accurate than for the helical case. After
storage time of $30$ $\mu s$ the dark center was completely filled
with light.} \label{fig:FlatRadialDistribution}
\end{figure}

  In figures \ref{fig:HelicalRadialDistribution} and
\ref{fig:FlatRadialDistribution} we present a quantitative
comparison between the calculated (solid lines) and
 the measured (symbols) cross-sections of the retrieved helical
 and flat phase beams, respectively, for different storage times. The total intensity of the stored beams
 decayed exponentially with a measured decay rate of
 $20,000$ $s^{-1}$. To facilitate the comparison between the cross-sections, both the
measured and the calculated curves are normalized to have same
total intensity.

 Since diffusion was strong in this experiment, we found that even without storing, the effect
of the diffusion on the slowed beam shape was already pronounced.
In fact, the dark center of the flat phase beam came out partly
filled when slowed (the slowing delay time was $\sim 50$ $\mu s$).
To account for this effect, the predicted curves of the retrieved cross-sections at different storage durations, shown in Fig. \ref{fig:HelicalRadialDistribution} and Fig. \ref{fig:FlatRadialDistribution}, were calculated by taking slowed beam profiles (helical or flat phase respectively) and propagating them forward in time, by the proper storage duration, using the three-dimensional diffusion propagator.
Alternatively, we found that we can obtain the shape of the slowed
beam by propagating the off-resonance beam (helical or flat phase)
by an effective storage duration using Eq. 1 (for the
helical beam) or Eq. 2 (for the flat phase beam). By further
propagating in time by the actual duration of the storage,
we were able to recover the predicted radial cross-section after
storage, also in reasonable agreement with the measured data.

 The best fit to the experiment was achieved when the
diffusion coefficient was $D=11$ $cm^2/s$. This is in good agreement
with an independent measurement of the diffusion coefficient we
performed using optical pumping, and also in reasonable agreement
with \cite{FranzDiffusion}.
From Fig. \ref{fig:HelicalRadialDistribution} and Fig. \ref{fig:FlatRadialDistribution} it is apparent that the theoretical prediction are able to capture the effect of the diffusion on the
retrieved signal. More importantly, it is apparent that the dark center of the stored helical
beam stays dark for times up to $110$ $\mu s$, in clear contrast to the flat phase beam.

 In summary, we have shown that it is possible to store and
retrieve an optical vortex. We found that due to its topological
nature, the dark center is stable to diffusion of the atoms. Since
diffusion is homogeneous and isotropic, atoms enter the dark center from all directions carrying a phase which is uniformly distributed over the unit circle. Upon reconstruction, they destructively
interfere thus maintaining the darkness. We compared this situation to
the case where the stored coherence is flat phased and with a dark
center. We found that in this case the dark center is illuminated
upon retrieval, already after $10$ $\mu s$, showing that atoms
indeed diffuse into the dark center. Propagating the initial
stored coherence field with the diffusion propagator, we were able
to predict the shape of the retrieved light and found good
agreement with the measured data.

 These results indicate a few interesting
directions that we wish to point out. Since the position of the
dark center is sensitive to phase gradients \cite{SternGerlachEIT}, one can make a grid
of helical beams, and measure the transverse magnetic field
gradient by comparing the stored image to the original one. Also, by
using the dark spot as a pixel, different two dimensional patterns
can be stored. Another possibility is to store more elaborate
types of topological structures such as knots \cite{KnotLight} or
skyrmions \cite{Skyrmions}. Although the latter requires a double
EIT scheme, which is experimentally more challenging, it holds the
prospect of introducing non-Abelian phases that might be used to
store quantum information in a novel way
\cite{doubleEITNonAbelian}.

This work was partially supported by the fund for encouragement of
research in the Technion and by the Minerva foundation.

\bibliographystyle{apsrev}

\end{document}